\documentclass[journal]{IEEEtran}
\ifCLASSINFOpdf
\else
  \usepackage{times}
  \usepackage[dvips]{graphicx}
  \usepackage{alltt}
  \usepackage{epsfig}
  \usepackage{amsfonts}
  \usepackage{color}

\fi
\begin{document}
%

\title{Transistor Switches using Active Piezoelectric Gate Barriers}

%
%
%

\author{Raj~K.~Jana, Arvind~Ajoy, Gregory~Snider, and~
        Debdeep~Jena
        \thanks{R. K. Jana, A. Ajoy, G. Snider, and D. Jena are with the Electrical Engineering Department, University of Notre Dame, Notre Dame,
IN, 46556 USA e-mail: rjana1@nd.edu; aajoy@nd.edu; snider.7@nd.edu; djena@nd.edu.}
}
\date{\today}

\maketitle

\begin{abstract}
This work explores the consequences of introducing a piezoelectric gate barrier in a normal field-effect transistor.  Because of the positive feedback of strain and piezoelectric charge, internal charge amplification occurs in such an electromechanical capacitor resulting in a negative capacitance.  The first consequence of this amplification is a boost in the on-current of the transistor.  As a second consequence, employing the Lagrangian method, we find that by using the negative capacitance of a highly compliant piezoelectric barrier, one can potentially reduce the subthreshold slope of a transistor below the room temperature Boltzmann limit of 60 mV/decade.  However, this may come at the cost of hysteretic behavior in the transfer characteristics.
\end{abstract}


\begin{IEEEkeywords}
Electrostriction, Electromechanical capacitor, Piezoelectric barrier, Negative capacitance, PiezoFET, Subthreshold slope
\end{IEEEkeywords}

%
\IEEEpeerreviewmaketitle

\section{Introduction}
\IEEEPARstart{S}CALING of the size of field-effect transistors (FETs) has improved their performance and integration densities in integrated circuits for over two decades. Most conventional transistors make use of a passive insulating barrier layer between the gate metal and the semiconductor channel to modulate the density of conduction channel electrons or holes. Because the intrinsic properties of a passive gate barrier do not change with the applied voltage, they impose certain fundamental limitations on the resulting device performance.

One such limitation is the subthreshold slope, i.e. the gate voltage required to change the drain current by an order of magnitude \cite{ChangIEEE10, TheisIEEE10}, given by $SS = m$ $\times$ 60 mV/decade at room temperature \cite{Salahuddinnano08, Taurbook}.  Here, $m = 1 + C_{sc}/C_{ins}$ is the `body factor', $C_{sc}$ is the semiconductor channel capacitance, and $C_{ins}$ is the gate insulator capacitance.  In a traditional FET switch with a passive gate dielectric such as SiO$_2$, $C_{ins} > 0$ and thus $m > 1$, which leads to $SS > 60$ mV/decade \cite{Taurbook}. This result, combined with circuit requirements for the on current $I_{on}$ and the on/off ratio $I_{on}/I_{off}$ establish a minimum supply voltage $V_{dd}$, which does not scale in direct proportion with feature size \cite{ChangIEEE10, TheisIEEE10, Meindl01}.  Scaling of $V_{dd}$ has hit a roadblock, giving rise to heat generation associated with the large power dissipation density in ICs \cite{ChangIEEE10, TheisIEEE10, HaenschIEEE06, Meindl01}, since the dissipated power is proportional to the square of the voltage, $P_{diss} \propto V^{2}_{dd}$ \cite{TheisIEEE10, JanaIEEE14}.  Many ideas based on alternate transport mechanisms in the semiconductor channel, such as interband tunneling, or impact ionization are being explored to lower $V_{dd}$.

An interesting alternative is to replace the passive gate barrier with an active one.  A first proposal of an active ferroelectric insulator \cite{Salahuddinnano08} predicts {\em internal voltage gain}: the voltage across the gate insulator layer is larger than the applied external gate voltage.  The origin of internal voltage gain is the collective alignment of the microscopic electric dipoles in the ferroelectric layer in response to the external electric field produced by the gate voltage.  The alignment of dipoles generates a voltage of its own, thus amplifying the voltage that makes it to the semiconductor channel.  Under appropriate bias conditions \cite{Salahuddinnano08}, the insulator capacitance provided by the ferroelectric is mathematically negative ($C_{ins} < 0$), causing $m = 1 + C_{sc}/C_{ins} <1$ and $SS < 60$ mV/decade.  Such an active-gate FET then will require a lower gate voltage to create the same charge as a conventional FET with passive gate dielectrics \cite{Salahuddinnano08}, thereby facilitating device scaling.

In this paper, we explore the device consequences of using a {\em piezoelectric} insulator as the active gate barrier in a transistor instead of the ferroelectric barrier.  Piezoelectric gate barriers are at the heart of commercially available III-nitride heterostructure transistors \cite{WangEDL10, HuDRC14}.  We first consider an active compliant piezoelectric layer as the insulator in a parallel plate capacitor.  We find that this simple electromechanical capacitor system exhibits a remarkably rich range of behavior.  We show that negative capacitance emerges as a natural response to applied voltage.  In this regime of negative capacitance, we show that we obtain a higher charge than in a corresponding capacitor with a passive dielectric.  Non-trivial capacitance-voltage behavior in such capacitors have also been reported experimentally \cite{Hemert11, ThenIEDM13}.  Next, we port the parallel-plate electromechanical capacitor to the gate capacitor of a FET.  We show how this piezoelectric gate stack enables a higher on-current than in a transistor with a passive dielectric due to internal charge amplification.  Finally, building upon our earlier proposals \cite{JanaPSS, JanaE3S, JanaIEDM14}, we discuss the possibility of using the negative capacitance regime of a highly compliant piezoelectric barrier to obtain sub-$60$ mV/decade switching in a transistor.
\section{Electromechanical capacitor}
\begin{figure}[htb]
\centering
\includegraphics[width=88 mm]{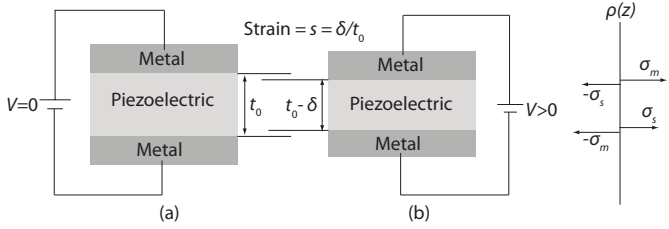}
\caption{ (a) Schematic cross section of a parallel-plate electromechanical capacitor with piezoelectric barrier layer of thickness $t_{0}$ at $V =$ 0 V, b) The layer thickness shrinks to $t_{0} - \delta$ when voltage $V$ is applied.  Sheet charge distribution $\rho(z)$ with $\pm \sigma_{m}$ on the metal plates and surface charges $\pm \sigma_{s}$ on the piezoelectric.}
\label{fig1}
\end{figure}

We begin by discussing the piezoelectric parallel-plate capacitor.  Consider the parallel plate capacitor of area $A$ shown in Fig. \ref{fig1} (a).  The equilibrium thickness $t_0$ of the piezoelectric insulator layer sandwiched between the metal plates changes to $t = t_0 - \delta$ when a voltage $V$ is applied on the plates, as shown in Fig. \ref{fig1} (b).  The strain is defined as $s = \delta / t_0$.  The equal and opposite sheet charges $\sigma_m$ that develop on the metal plates set up an attractive force between them, which strains the insulator.  This effect, called electrostriction, is the electric-field induced reduction of the thickness of a material; it occurs in {\em all} insulators, whether or not the layer is piezoelectric.  However if the insulator is piezoelectric, the strain {\em amplifies} the surface charge of the insulator.  This mechanism sets up a positive feedback between the thickness and the electric field, and is responsible for the appearance of negative capacitance.  To find the capacitance in the presence of such electromechanical coupling, one must first find the net metal charge $\sigma_m$ as a function of the external (battery) voltage, and then take its derivative.  This requires us to identify the surface charges $\sigma_s$ that develop at the surface of the insulator.  The resulting electric field profile is constant, equal to $E = V/t$, and the voltage drops linearly across the insulator.

Maxwell's boundary conditions across the metal-insulator interface requires the normal components of the displacement vector to obey $D_{d} - D_{m} = \sigma_m$. $D_d$ is the displacement field in the dielectric related to the surface charges $\sigma_s$ by $D_d = \epsilon_0 E + \sigma_s$, where $\sigma_s = ( \epsilon_d - \epsilon_0 ) E + e_{33}s + \sigma_{sp}$ or $D_{d} = \epsilon_d E + e_{33} s + \sigma_{sp}$. Here $\epsilon_d = \epsilon_0 (1 + \chi_d)$ is the net dielectric constant of the piezoelectric layer, and $\chi_d$ is its electric susceptibility.  The electric field is $E= V/t$, where $t = t_{0}(1-s)$ is the thickness of the {\em strained} insulator layer.  We explicitly allow for both piezoelectric and spontaneous polarization for an active dielectric material.  The strain-induced piezoelectric contribution to the charge (to linear order) is $e_{33} s$, where $e_{33}$ is the piezoelectric coefficient in units of C/m$^2$ and $s = \delta / t_0$ is the strain along the field.  The charge due to spontaneous polarization is $\sigma_{sp}$, also in units of C/m$^2$.  Inside the metal, $D_m = 0$.  Therefore, we obtain the relation

\begin{equation}
\sigma_m = \epsilon_d \frac{V}{t_{0} (1 - s)} + e_{33} s + \sigma_{sp}.
\label{metalcharge1}
\end{equation}

This relation illustrates how the strain $s$ explicitly enters the {\em electrostatic} relation between the metal charge and the voltage across the capacitor.  If one neglects the spontaneous polarization ($\sigma_{sp} \rightarrow 0$), piezoelectric effect ($e_{33} \rightarrow 0$) and strain ($s\rightarrow 0$), we get $\sigma_m = C_0 V$, (with $C_0 = \epsilon_d / t_0$), the standard textbook formula of a parallel plate capacitor.  However, we note that one can turn off the spontaneous and piezoelectric polarization by choice of material, and yet the factor $(1-s)$ in the denominator will persist: this is the electrostriction term.

\begin{figure*}[t]
\centering
\includegraphics[width=5 in]{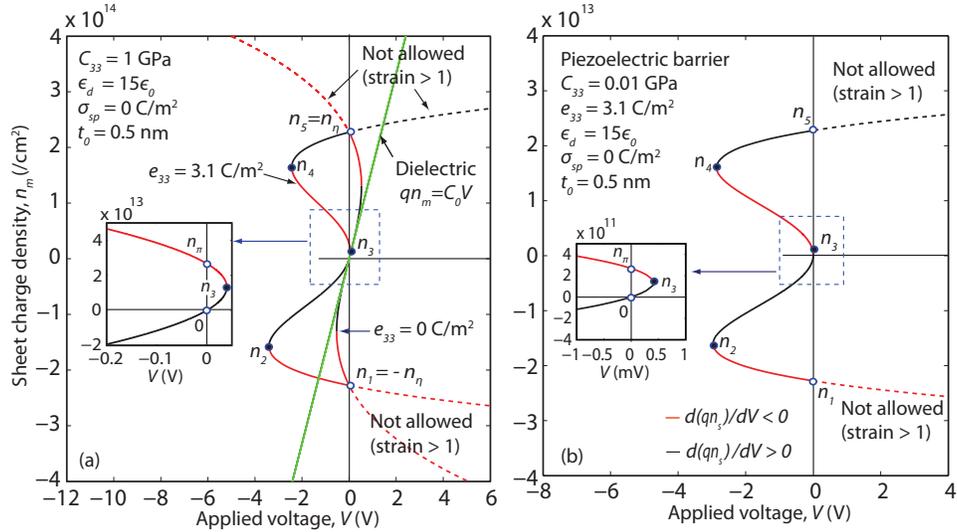}
\caption { a) Charge-voltage ($qn_{m}-V$) characteristic of the electromechanical capacitor.  Various charge states such as the positive capacitance segments $[n_2, n_3]$, $[n_4, n_5]$ where the slopes $C_{PE} = d(qn_m)/dV > 0$ are positive, and negative capacitance segments $[n_1, n_2], [n_3, n_4]$ where $C_{PE} = d(qn_m)/dV < 0$ are shown, b) The characteristics of a piezoelectric capacitor with a lower stiffness and more compliant barrier with $C_{33}=$ 0.01 GPa makes negative capacitance accessible at a lower charge $\sim 10^{11}$ cm$^{-2}$, as shown in the inset.}
\label{fig2}
\end{figure*}

The mechanical pressure $P$ experienced by the insulator is the electrical force $F$ per unit area $A$.  It is thus related to the metal charge \cite{Pelrine, Rivera} via $P = F/A =  \sigma_m^2 / \epsilon_d $.  To linear order, the pressure depends on the strain via the stiffness coefficient $P = C_{33} s$, where $C_{33}$ is in units of N/m$^2$, or Pascals.  Thus, we obtain the strain as a function of the metal charge: $s = \sigma_m^2/ \epsilon_d C_{33}$.  Substituting in Eq. \ref{metalcharge1} and rearranging, we have the desired relation between the metal charge in response to an applied voltage:

\begin{equation}
C_{0} V = \sigma_m - \sigma_{sp} + \frac{ ( \sigma_{sp} - e_{33} ) }{ \epsilon_d C_{33} } \sigma_m^2 - \frac{1}{\epsilon_d C_{33}} \sigma_m^3 + \frac{ e_{33} }{ (\epsilon_d C_{33})^2 } \sigma_m^4.
\label{metalcharge2}
\end{equation}

The right hand side is a fourth order polynomial in $\sigma_m$, and captures the electromechanical coupling physics.  Let us explore its consequences.  The sheet charge on the metal $n_m = \sigma_m/q$ from Eq. \ref{metalcharge2} is plotted as a function of the applied voltage $V$ in Fig. \ref{fig2} for different sets of material parameters.  For example, $e_{33}  = 3.1$  C/m$^2$,  $\epsilon_d  = 15  \epsilon_0$ correspond to the piezoelectric material Sc$_x$Al$_{1-x}$N \cite{AkiyamaAPL13, Tasna´diPRL10}.  The value of $C_{33}$ is allowed to vary arbitrarily in order to investigate the range of behavior of the piezoelectric capacitor.  We also assume $\sigma_{sp} = 0$. A non-zero value of $\sigma_{sp}$ merely causes a horizontal shift of the $\sigma_m - V$ curve (see Supporting document).  The physics of the piezoelectric capacitor with $\sigma_{sp} =0$ becomes apparent by factoring Eq. \ref{metalcharge2} into
\begin{equation}
V = \frac{q}{C_0} \left( n_m \right) \left( 1 - \frac{n_m}{n_{\eta}} \right)
                 \left( 1 + \frac{n_m}{n_{\eta}} \right)
                 \left( 1 - \frac{n_m}{n_{\pi}} \right),
\label{metalchargePZ}
\end{equation}
where $qn_m = \sigma_m$, $q n_{\pi} =  \epsilon_d C_{33}  / e_{33}$, and $q  n_{\eta} = \sqrt{\epsilon_d C_{33}}$.  Setting $V=0$ in Eq. \ref{metalchargePZ}, we obtain four {\em real} roots $n_{m0} =$ $0$, $+ n_{\eta}$, $- n_{\eta}$  and  $+ n_{\pi}$.  For a rigid ($C_{33} \rightarrow \infty$), non-piezoelectric ($e_{33}=0$) insulator, $n_{\eta},  n_{\pi}  \rightarrow  \infty$, whereupon we recover $\sigma_m = qn_{m}  = C_{0} V$, and the metal charge is a linear function of voltage as shown in the green line in Fig. \ref{fig2}(a).

On the other hand, for a compliant non-piezoelectric insulator, $C_{33} > 0$, and Eq. \ref{metalchargePZ} reduces to a cubic equation with roots $0, \pm n_{\eta}$ at $V=0$. This is in fact a prototypical description of a nano-electromechanical switch \cite{Masuduzzaman14}.  The two additional roots $\pm n_{\eta}$ make the dependence of $\sigma_m$ on $V$ {\em nonlinear} with two additional zero crossings.  Multiple zero crossings of the $qn_m - V$ curve mathematically guarantees that there must be regions of negative slope $d(qn_m)/dV < 0$.  This is shown in red in the flipped S-shaped curve of Fig. \ref{fig2}(a), where $C_{33} = 1$ GPa is assumed.  In these regions, the electromechanical capacitor has a negative capacitance.

\begin{figure}[t]
\centering
\includegraphics[width=88 mm]{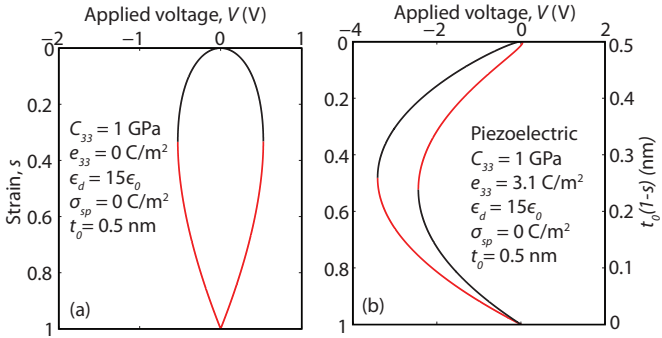}
\caption{ Strain as a function of voltage in a) a non-piezoelectric  insulator layer with $C_{33}=$ 1 GPa, $e_{33}=$ 0 C/m$^{2}$, $t_{0}=$ 0.5 nm, and b) in a piezoelectric insulator layer with $C_{33}=$ 1 GPa, $e_{33}=$ 3.1 C/m$^{2}$.  Strain $s< 1$ is physically accessible in solid state, where the remaining layer thickness, $t_0 (1 - s) > $ 0.  Red (black) represents strain corresponding to the negative (positive) capacitance charge states.}
\label{fig3}
\end{figure}

However, the negative capacitance corresponds to very high values of charge density($> 10^{14}$ cm$^{-2}$) and strain $s$ ($>$ 0.3), as shown in Fig. \ref{fig3}(a).  Though rapid progress is being made in the solid-state electrostatic gating of ever increasing carrier densities in semiconductors \cite{VermaAPL14}, methods to reduce the charge and strain are desirable.  Now consider the piezoelectric insulator where $C_{33} > 0$ and $e_{33} > 0$. Notice that $n_{\eta}$ is independent of $e_{33}$.  The root $n_{\pi}$ depends on $e_{33}$ and its location determines the shape of the  $qn_m - V$ curve.  If $e_{33} > \sqrt{\epsilon_d C_{33}}$, then $0 <  n_{\pi} < n_{\eta}$ and negative capacitance appears in the two charge segments $[n_1=-n_{\eta}, n_2]$ and $[n_3,n_4]$ shown in red in Fig. \ref{fig2}.  It is important to realize that piezoelectricity lowers both the charge density ($\sim 10^{13}$ cm$^{-2}$) [inset of Fig. \ref{fig2}(a)] and strain $ < 0.01$ [Fig. \ref{fig3}(b)] at which negative capacitance appears, compared to electrostriction alone.
For vanishingly small voltages around zero, the piezoelectric capacitor behaves exactly like a parallel plate capacitor - a straight line.  But the additional benefit of the above coupling is the increased charge density compared to a passive dielectric due to the piezoelectric amplification - this effect will boost the on-state current in a transistor.  From Eq. \ref{metalchargePZ}, the piezoelectric amplification is $n_m - \frac{C_0 V}{q} \approx \frac{n_m^2}{n_\pi} + ...$ to leading order.  Finally, if we use a highly compliant piezoelectric, for example with $C_{33} = 0.01$ GPa, negative capacitance can be accessed at very low charge density $\sim 10^{11}$ cm$^{-2}$, as indicated in Fig. \ref{fig2}(b).  These highly compliant piezoelectrics can potentially enable the design of transistors with steep sub-threshold behavior, but require new materials as will be described later.  We also remark here that Pauli's exclusion principle of solid matter and quantum compressibility restricts $s < 1$. Therefore for piezoelectric insulators, the metal charge will be restricted to $- n_{\eta} < n_m < + n_{\eta} $.  It may be possible to go beyond these restrictions ($s > 1$, shown as dashed lines in Fig. \ref{fig2}) in gaseous plasmas where charged ion plate `electrodes' can pass through each other.  But we do not pursue that line of analysis here, by restricting the discussion to solid metals and dielectrics.

\section{Transistor with a Piezoelectric Barrier}
\begin{figure}[htb]
\centering
\includegraphics[width=88 mm]{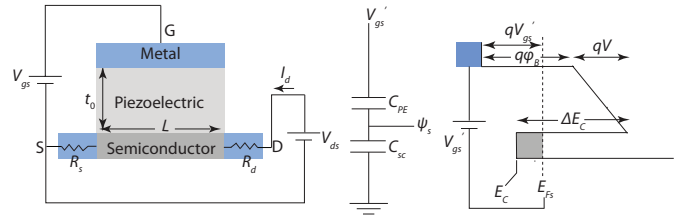}
\caption{ Schematic cross section of a transistor (``piezoFET") with a piezoelectric gate barrier, semiconductor channel such as Si, GaN or 2D material, MoS$_{2}$, and source and drain ohmic contacts.  The gate capacitance circuit is a series combination of the piezoelectric capacitance $C_{PE}$ and the semiconductor capacitance $C_{sc}$.  Here intrinsic gate voltage $V_{gs}' = V_{gs} - I_d Rs$, and intrinsic drain voltage $V_{ds}' = V_{ds} - I_d(R_s + R_d)$, where $R_s$ and $R_d$ are the source and drain contact resistances.  The energy band diagram is shown for the metal-piezoelectric-semiconductor stack of the transistor. $\psi_s = V_{gs}' - V = (E_{Fs} - E_C)/q$ is the surface potential.}
\label{ballistic_fet}
\end{figure}

\begin{figure*}[htb]
\centering
\includegraphics[width=7 in]{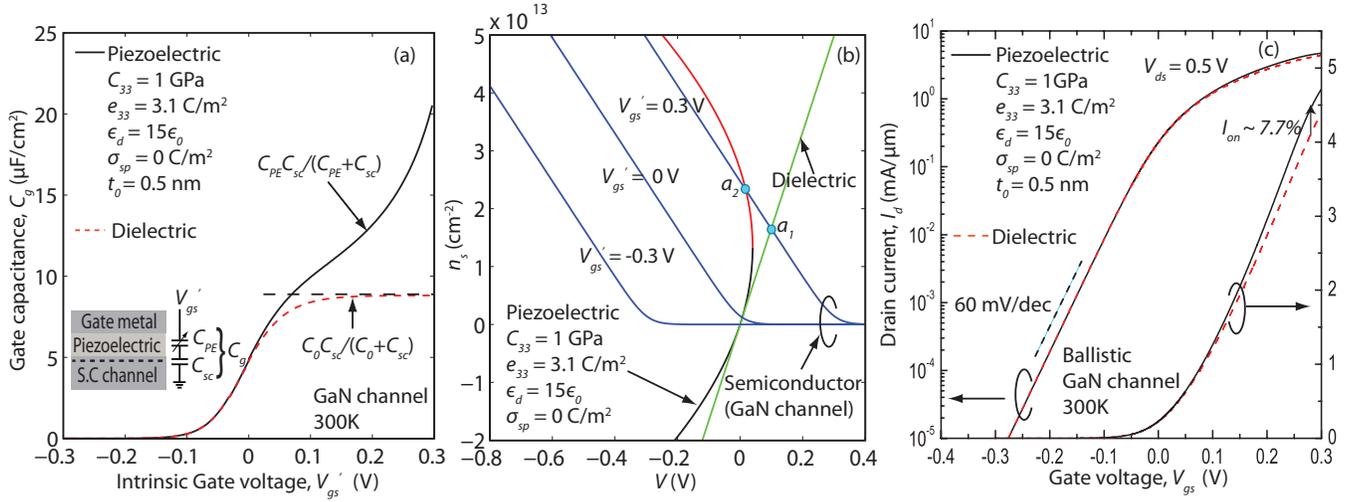}
\caption{ a) Gate capacitance $C_g$ versus $V_{gs}'$ for transistors with piezoelectric (solid line) and dielectric (dashed line) insulators, b) Graphical load line analysis to obtain sheet carrier density $n_{s}$ for different $V_{gs}'$.  Blue curves show the semiconductor charge for different $V_{gs}'$, green shows the metal charge in the case of a passive dielectric, whereas red(black) shows the metal charge in the negative(positive) capacitance regimes of the piezoelectric capacitor.  Intersections $a_1$ and $a_2$ of the above characteristics define the operating points of the system, c) Transfer curve depicts the drain current $I_d$ versus $V_{gs}$ at drain voltage $V_{ds} =$ 0.5 V for GaN transistors with piezoelectric (solid line) and dielectric barriers (dashed line).}
\label{fig5}
\end{figure*}

We now explore how the presence of the piezoelectric capacitor in the gate of a transistor with a semiconductor channel changes the traditional characteristics.  The semiconductor channel could be formed of a gapped 2-dimensional crystal such as MoS$_2$, or a 3-D crystal semiconductor such as Si or GaN.  The semiconductor is characterized by the valley degeneracy $g_{v}$ of the conduction (or valence) band.  We assume the energy dispersion of each valley to be the same, characterized by an effective mass $m^{\star}$ and spin degeneracy $g_s=2$.  Carrier transport in the semiconductor channel is assumed to be 2-dimensional - which holds both for monolayer 2D crystals and in field-effect transistors made of 3D semiconductors, where transport occurs in a quasi-2D electron/holes gas.  The occupation of multiple 2D subbands can then be treated as individual 2D channels - we consider a single subband model.

\begin{figure*}[htb]
\centering
\includegraphics[width=5 in]{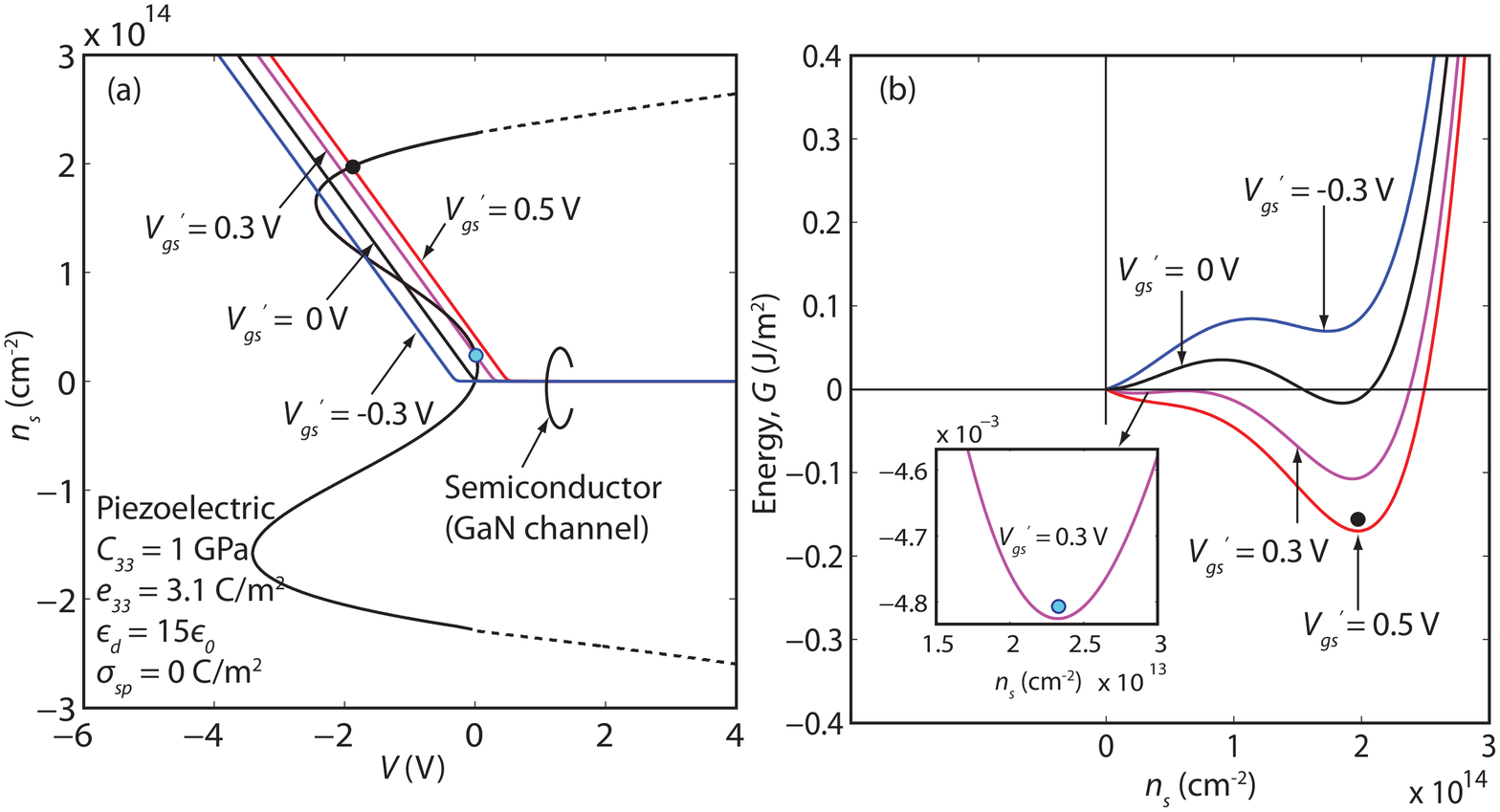}
\caption { a) Load line analysis showing multiple intersections of the piezoelectric and semiconductor characteristics for different $V_{gs}'$, b) Free-energy landscape of the piezoelectric-semiconductor stack at various $V_{gs}'$.  Blue and black dots show stable operating points.}
\label{fig6}
\end{figure*}

\begin{figure*}[htb]
\centering
\includegraphics[width=7 in]{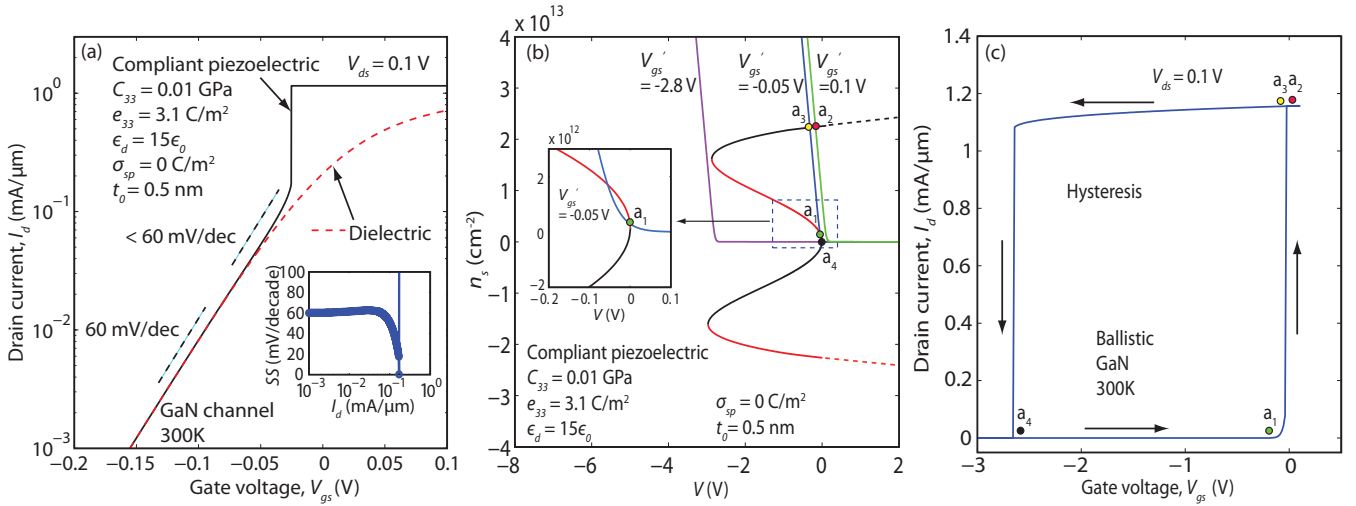}
\caption{a) $I_d$ versus $V_{gs}$ curves at $V_{ds} =$ 0.1 V for a GaN channel piezoFET with a compliant piezoelectric with $C_{33} =$ 0.01 GPa.  A boost in the on-current and sub-$60$ mV/dec SS (inset) are obtained as compared to a passive gate dielectric. b) Load line characteristics to explain the hysteresis with gate bias voltages $V_{gs}'$, c) The calculated hysteresis in the transfer curve $I_d$ versus $V_{gs}$ for forward and reverse sweeps is shown for a GaN channel piezoFET.}
\label{fig7}
\end{figure*}

The semiconductor channel of length $L$ and width $W$ is assumed to be connected to very low-resistance ohmic contacts at the source and drain, as shown in Fig. \ref{ballistic_fet}.  The energy-band diagram in Fig. \ref{ballistic_fet} shows the potential barrier controlled by the voltage on the gate metal.  Electrical charge neutrality requires $\sigma_{m} = q n_{s}$, where $n_s$ is the mobile carrier sheet density at the `top-of-the-barrier' in the energy band diagram along the length of the channel.  The energy band diagram from the metal to the semiconductor requires $q\phi_B + q V - \Delta E_c + (E_{Fs} - E_c) = q V_{gs}'$.  By suitable choice of materials, we assume that $q \phi_B = \Delta E_c$; if this is not the case, the difference can be absorbed in a shift of threshold voltage.  When no drain voltage is applied, carriers in the semiconductor are in thermal equilibrium with the source and drain reservoirs, which for a parabolic 2D bandstructure means $qn_{s} = C_{sc} V_{th} \ln \left( 1 + \exp [ (E_{Fs} - E_{c})/kT ] \right)$, or $E_{Fs} - E_c = kT \ln \left( \exp[ q n_s / C_{sc} V_{th} ] - 1 \right)$, where $C_{sc}=q^2 g_s g_v m^{\star} / 2 \pi \hbar^2$ is the density of states semiconductor capacitance, and the thermal voltage $V_{th} = kT / q$.  From the energy band diagram in Fig. \ref{ballistic_fet}, the relation between the applied gate voltage $V_{gs}'$ and the voltage drop $V$ across the piezoelectric insulator is $qV_{gs}' = qV + (E_{Fs} - E_C )$. Here, $(E_{Fs} - E_c)/q = (V_{gs}' - V) = \psi_s$ is the surface potential.  Using the carrier density expression and Eq. \ref{metalcharge2}, the gate-induced charge $qn_s$ in the semiconductor channel is self-consistently calculated.  Finally, using this new dependence of charge on the voltages and the piezoelectric coefficients, the current-voltage characteristics of the piezoFET are obtained from the ballistic transport model \cite{NatoriJAP94} incorporating the quantum contact resistances of 0.026 k$\Omega$.$\mu$m \cite{JenaNature14} at the source and drain ends.

Fig. \ref{fig5} shows the gate capacitance $C_g = d (qn_s)/dV_{gs}'$, and device characteristics ($I-V$) of a ballistic piezoFET with a GaN channel ($m^{\star} =$ 0.2 $m_0$, $g_v =$ 1 \cite{Woodbook}).  Fig. \ref{fig5}(a) shows that a higher gate $C_g$ is obtained in the piezoFET (solid line), as compared to a FET with a passive gate (dashed line).  The higher $C_g$ is due to the negative capacitance resulting from piezoelectric charge amplification: $C_{PE}C_{sc}/(C_{PE} + C_{sc}) > C_0C_{sc}/(C_0 + C_{sc})$ when $C_{PE} < 0$.  Fig. \ref{fig5}(b) depicts the solution of the piezoelectric and semiconductor charge equations graphically, following the load-line approach (see \cite{ChenTED11}).  The blue lines depict charge in the semiconductor channel, and the green, black, and red lines depict the charge drawn into the metal from the battery.  They must be equal to maintain global charge neutrality, meaning the locus of intersections are the operating points of the device.  The green line is the charge on the metal for a traditional passive gate dielectric, and the red/black lines for a piezoelectric gate.  When the transistor is on ($V_{gs}' \sim$ + 0.3 V), an increase in the charge at point $a_2$ in Fig. \ref{fig5}(b) is seen for the piezoelectric compared to point $a_1$ for a passive dielectric.  This increased charge boosts the on-current as depicted in Fig. \ref{fig5}(c), consequently improving the $I_{on}/I_{off}$ ratio.  This sort of piezoelectric amplification is an interesting method to boost the on-current in {\em any} transistor.  Since much of the high-performance characteristics such as gain and cutoff frequencies depend on $I_{on}$, corresponding boosts can be expected in these parameters.  This may be specially useful for boosting the current in FETs made of relatively low mobility channel materials.  Note however in Fig. \ref{fig5}(c) that this device still has a SS of $60$ mV/decade. This is because the negative capacitance regime is only accessible for charge densities $\geq 1.5 \times 10^{13}$ cm$^{-2}$: at this high level of charge, the transistor is in its on-state, rather than in the sub-threshold regime.

Because the charge-voltage characteristic of the piezoelectric capacitor is highly non-linear, it can have multiple intersections with the semiconductor load line.  Ref. \cite{CherryPhilMag1951} develops a systematic procedure to understand such non-linear systems based on the Euler-Lagrange equations of motion (see supporting document for details).  For this analysis, we define a free-energy $G$ in units of J/m$^2$ for the piezoelectric-semiconductor stack: $G(\sigma_m, V_{gs}') = \int V d\sigma_m + \int \psi_s d\sigma_m - \sigma_m V_{gs}'$, where $V$ is the voltage drop across the gate insulator and $\psi_s$ is the surface potential.  Minima in this free-energy landscape correspond to stable charge solutions of the non-linear system. If there are multiple minima, the actual solution $\sigma_m = q n_s$ depends on the previous state, or the {\em history} of the system.

For example, Fig. \ref{fig6}(a) shows the load lines and the corresponding evolution of the free-energy landscape for different $V_{gs}'$ are shown in Fig \ref{fig6}(b).  The shape of the energy landscape changes with the applied voltage.  There are two energy minima in the range $-0.35 < V_{gs}' < 0.35$ V, and a single minimum otherwise.  Let us assume that there is no charge to begin with on the capacitor, and ramp the gate from a negative to a positive voltage.  Until around $V_{gs}'=$ + 0.35 V, the system remains in the minimum corresponding to the lower charge state ($\sim 2.3 \times 10^{13}$/cm$^2$) shown as a blue dot in Fig. \ref{fig6}(a) and the inset of Fig. \ref{fig6}(b).  But when $V_{gs}' >  0.35$ V it is driven into the higher charge state shown as a black dot.  Thus, provided $V_{gs}' < 0.35$ V, the transistor displays no hysteresis in its $I-V$ characteristics.

It is pertinent here to note an important difference in the nature of negative capacitance of the piezoelectric and ferroelectric insulators. The ferroelectric capacitor possesses negative capacitance at zero charge, whereas the capacitance of the piezoelectric capacitor is positive at zero charge. This property of the ferroelectric capacitor is exploited in achieving $SS < 60$ mV/decade, since the semiconductor load line can intersect the negative capacitance regime of the ferroelectric characteristic at the very low charge densities corresponding to subthreshold operation of the transistor.  Can a similar negative capacitance be obtained in the SS regime ($V_{gs}'< 0$ V) using piezoelectric gates?  We explore this by tuning the piezoelectric material properties.

We find that if a lower stiffness, highly compliant piezoelectric barrier with $C_{33} \sim $ 0.01 GPa is used, it can enable the reduction of the subthreshold slope below $60$ mV/decade and also boost the on-current.  This is shown in Fig. \ref{fig7}(a).  Here negative capacitance is accessed in the subthreshold region, shown by the operating point $a_1$ in the load line characteristics at $V_{gs}'=$ -0.05 V shown in Fig. \ref{fig7}(b).  The on-state operation of this transistor corresponds to the higher charge state determined by the operating point $a_2$ in the load line characteristics at $V_{gs}'=$ 0.1 V.  However, this also results in hysteresis in the transistor characteristics with $V_{gs}'$ sweep, as shown in the $I_d-V_{gs}$ characteristics in Fig. \ref{fig7}(c) which is calculated using the Lagrangian method.  Hysteresis is undesirable in purely switching applications, but desirable for memory.  Further, the strain in the higher charge states $a_2, a_3$ is very close to $100\%$, which is not feasible in realistic materials.  If suitable new piezoelectric materials with ultra-low $C_{33}$ and high $e_{33}$ could be developed (see Supporting Information for various piezoelectrics with different $C_{33}$ and $e_{33}$), sub-$60$ mV/decade switching can be achieved with hysteresis with suitable choice of semiconductors.  Investigation of other transistor designs incorporating the piezoelectric barrier, such as the quantum metal transistor \cite{FrankTED14} to eliminate the hysteresis and reduce strain could be the focus of future work.

\section{Conclusion}
We also emphasize that we have assumed linear piezoelectric parameters in this work to keep the model simple and yet capture the new physics.  The non-linear material response needs to be explored in future.  To conclude, the behavior of transistor switches using active piezoelectric gate barriers was explored.  Because of electrostriction and piezoelectricity, negative capacitance is predicted to appear in a piezoelectric capacitor.  Using this negative capacitance and a ballistic transport model, we predict that compliant piezoelectric barriers can boost the gate capacitance and increase the on-currents of transistors.  Also, steep switching with sub-60 mV/decade subthreshold slope is predicted when the negative capacitance of the piezoelectric barrier is accessed in the off-state operation of the transistor, and this steep behavior is predicted to be assisted by hysteresis based on the Lagrangian method of stability of the transistor system.



%



\section*{Acknowledgment}
This work was supported by the Center for Low Energy Systems Technology (LEAST), one of six centers of STARnet, the Semiconductor Research Corporation (SRC) program sponsored by MARCO and DARPA.


\ifCLASSOPTIONcaptionsoff
  \newpage
\fi



%


\setcounter{equation}{0}
\setcounter{figure}{0}
\setcounter{table}{0}
\setcounter{page}{1}
\setcounter{section}{0}
\renewcommand{\theequation}{S\arabic{equation}}
\renewcommand{\thefigure}{S\arabic{figure}}
\renewcommand{\thesection}{S\arabic{section}}

\begin{center}
\onecolumn
\textbf{\Large Supporting Information \\ Transistor Switches using Active Piezoelectric Gate Barriers}
\end{center}


\section{$\sigma_m-V$ relation of piezoelectric capacitor with $\sigma_{sp}$}

Following the same notation as the main text, the 4$^{th}$ order charge versus voltage $\sigma_m-V$ relation of an electromechanical capacitor is
\begin{equation}
C_{0} V = \sigma_m - \sigma_{sp} + \frac{ ( \sigma_{sp} - e_{33} ) }{ \epsilon_d C_{33} } \sigma_m^2 - \frac{1}{\epsilon_d C_{33}} \sigma_m^3 + \frac{ e_{33} }{ (\epsilon_d C_{33})^2 } \sigma_m^4.
\label{metalcharge_2}
\end{equation}

\begin{figure}[htb]
\centering
\includegraphics[width=60 mm]{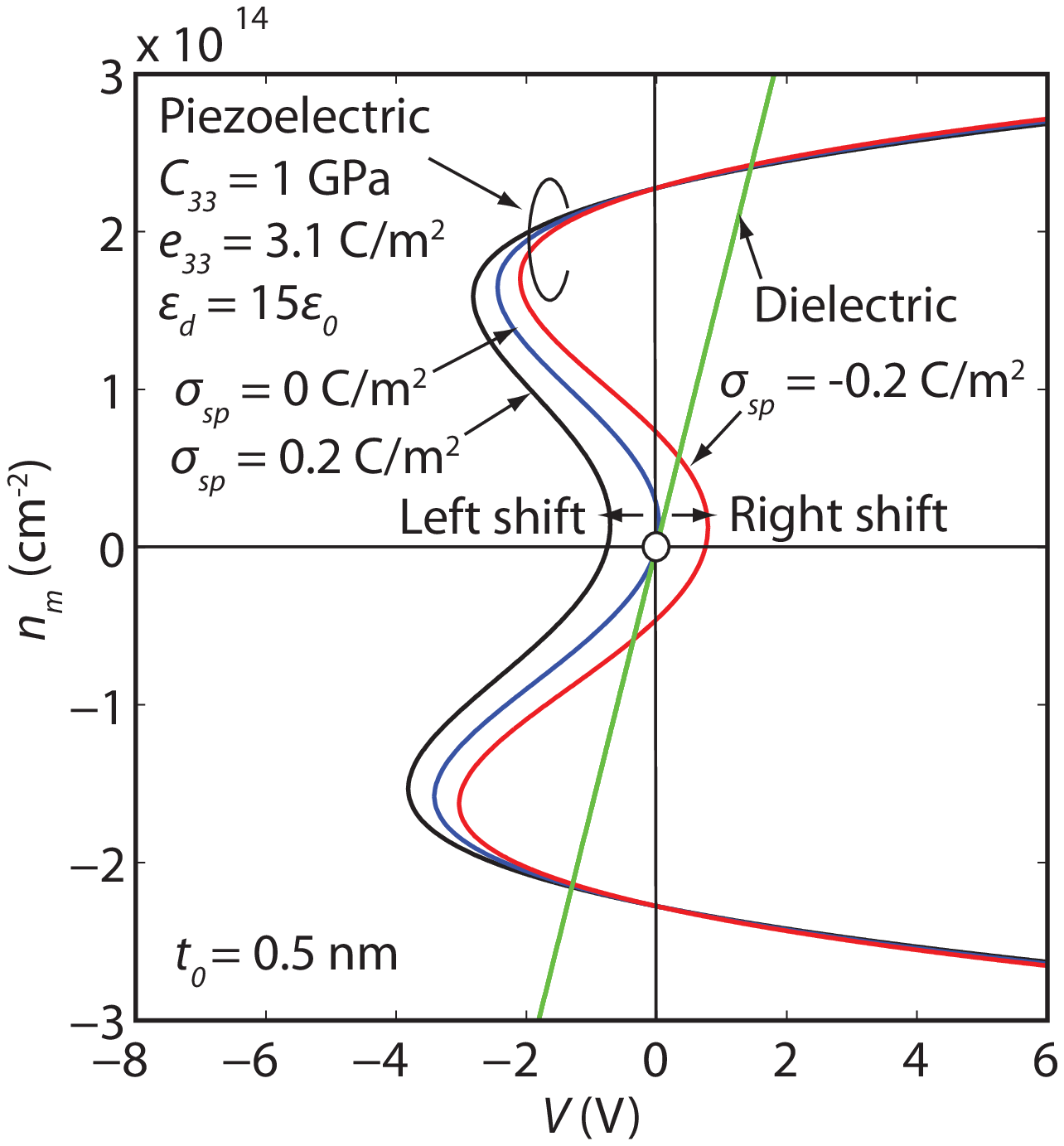}
\caption{ Charge-voltage ($qn_{m}-V$) characteristics of an electromechanical capacitor with piezoelectric barrier.  Color lines (black, blue and red) show the characteristics of capacitors for $\sigma_{sp} =$ 0.2 C/m$^{2}$, 0 C/m$^{2}$, and -0.2 C/m$^{2}$.  $\sigma_{sp} \neq$ 0 C/m$^{2}$ leads to  horizontal left and right shifts of $qn_{m}-V$ curve.}
\label{fig8}
\end{figure}

The right side is a fourth order polynomial in $\sigma_m$, and captures the electromechanical coupling physics.  Writing $\sigma_m = q n_m$ where $n_m$ is the sheet charge density on the metal and $\sigma_{sp} = q n_{sp} \neq 0$, the polynomial factorizes to

\begin{equation}
C_{0} V = q n_{sp} ( 1 - \frac{ n_m }{ n_{\alpha}^{+} } )( \frac{ n_m }{ n_{\alpha}^{-} } - 1 )( 1 - \frac{ n_m }{ n_{\eta} } )( 1 + \frac{ n_m }{ n_{\eta} } ) ,
\label{metalcharge_SP}
\end{equation}

where $n_{\alpha}^{+} = (n_{\pi} + \sqrt{ n_{\pi}^2 - 4 n_{sp} n_{\pi} })/2$, $n_{\alpha}^{-} = (n_{\pi} - \sqrt{ n_{\pi}^2 - 4 n_{sp} n_{\pi} })/2$, $q n_{\pi} = \epsilon_d C_{33} / e_{33}$, and $q n_{\eta} = \sqrt{\epsilon_d C_{33}}$.  The four roots are characteristic sheet densities determined uniquely by the electromechanical coefficients and the spontaneous polarization of the dielectric material.  If $n_{\pi} > 4 n_{sp}$ which is met if $C_{33} > 4 n_{sp} e_{33} / \epsilon_d$, then all four roots are real.  The effect of spontaneous polarization $\sigma_{sp}\neq$ 0 C/m$^{2}$ is a voltage offset, which leads to left and right shifts of the $qn_m-V$ characteristics (black and red curves) with respect to the characteristic (blue curve) with $\sigma_{sp}=$ 0 C/m$^{2}$, as shown in Fig. \ref{fig8}.  These shifts will move the threshold voltages of a corresponding transistor, and will also locally change the slopes of the charge-voltage characteristics.

\section{Ballistic FET I-V model}

\begin{figure*}[htb]
\centering
\includegraphics[width=120 mm]{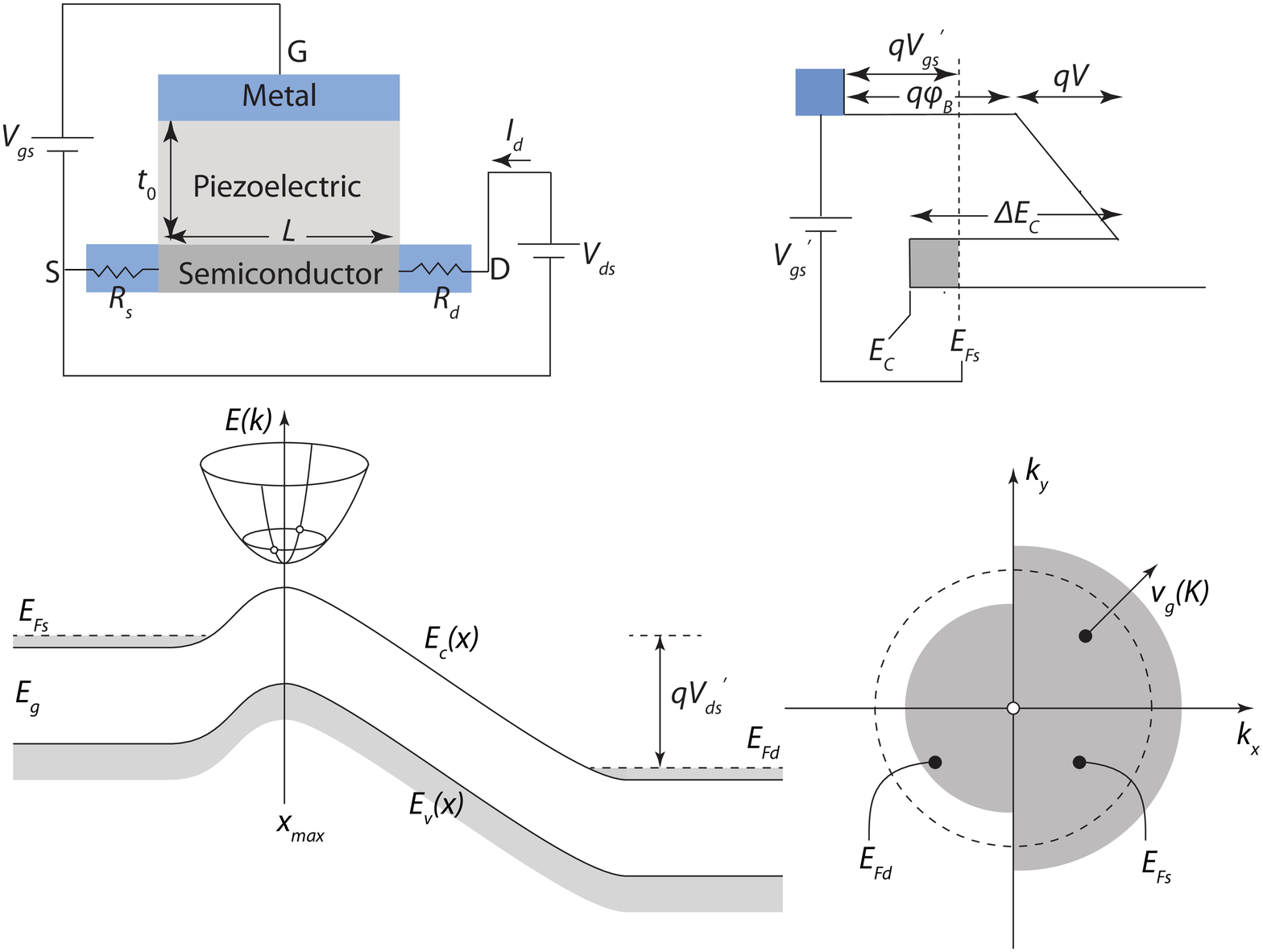}
\caption{ Schematic cross section of a transistor (``piezoFET") with a piezoelectric gate barrier, semiconductor channel such as Si, GaN or 2D material, MoS$_{2}$ and source and drain ohmic contacts.  Energy band diagram for the metal-piezoelectric-semiconductor stack of the transistor.  For transistor operation, we use ballistic transport model to calculate the transistor characteristics.  The mobile sheet carrier density at the `top-of-the-barrier' in energy band diagram is controlled by $V_{gs}$ through the piezoelectric gate barrier, and source and drain Fermi levels $E_{Fs}$ \& $E_{Fd}$ are separated by $V_{ds}$.  Dashed circle shows the $k$-space for carrier distributions at $V_{ds}=0$ and grey circle shows the $k$-space for carrier distributions at applied $V_{ds}^{'}$ i.e. $E_{Fs} - E_{Fd} = q V_{ds}^{'}$.  $v_{g}(k)$ is the group velocity of carriers.  Here, $V_{ds}^{'}=V_{ds}-I_d (R_s+R_d)$, where $R_s$ and $R_d$ are the source and drain contact resistances.}
\label{ballistic_fet1}
\end{figure*}

When no drain voltage is applied, carriers in the semiconductor are in thermal equilibrium with the source and drain reservoirs, which for a parabolic 2D bandstructure means $qn_{s} = C_{sc} V_{th} \ln \left( 1 + \exp [ (E_{Fs} - E_{c})/kT ] \right)$.  From the energy band diagram of metal-piezoelectric-semiconductor stack of a transistor (Fig. \ref{ballistic_fet1}), the voltage division $qV_{gs}^{'} = qV + (E_{Fs}^{0} - E_{c})$ translates to the dimensionless equation,

\begin{equation}
e^{\frac{ V_{gs} }{ V_{th} } }  = e^{ \frac{ q  F(n_s)}{ C_{0} V_{th} } } ( e^{ \frac{ q n_s }{ C_q V_{th}} } - 1 ),
\label{piezoSCcharge}
\end{equation}

where we have defined $F(n_s) = n_{sp} ( 1 - \frac{ n_s }{ n_{\alpha}^{+} } )( \frac{ n_s }{ n_{\alpha}^{-} } - 1 )( 1 - \frac{ n_s }{ n_{\eta} } )( 1 + \frac{ n_s }{ n_{\eta} } )$.  This equation must be solved to find the semiconductor charge $n_s$ at a gate voltage $V_{gs}$.  If the dielectric is piezoelectric but does not have spontaneous polarization, one must replace $F(n_s)$ by $F_{pz}(n_s) = n_s ( 1 - \frac{ n_s }{ n_{\eta} } )( 1 + \frac{ n_s }{ n_{\eta} } )( 1 - \frac{ n_s }{ n_{\pi} } )$ to find the semiconductor charge in response to the gate voltage.  This is the major change to a standard Natori-type ballistic FET model \cite{NatoriJAP_94} brought about by the piezoelectric gate barrier.

When a drain voltage $V_{ds}$ is applied, the carrier distribution in the `top-of-the-barrier' point $x_{max}$ in the energy band diagram is split in two.  In the ballistic limit of transport, the right-going carriers are in equilibrium with the source reservoir of Fermi energy $E_{Fs}$, whereas the left-going carriers are in equilibrium with the drain reservoir $E_{Fd}$, and they are out of equilibrium by $E_{Fs} - E_{Fd} = q V_{ds}^{'}$.  Note that $E_{Fs}^{0} \neq E_{Fs}$; the application of a drain bias causes a rearrangement of the carrier distribution in the semiconductor channel.  However, with good electrostatic design, one can ensure that the {\em net} carrier density at $x_{max}$ is the same as when $V_{ds}^{'} = 0$.  The carrier distribution in the $k-$space is depicted in Fig. \ref{ballistic_fet1}; the dashed circle is the distribution for $V_{ds}^{'}=0$, and the gray half-circles are the result of application of a drain bias, both in the $T \rightarrow 0$ K limit.  Defining $\eta_s = (E_{Fs} - E_c)/k T$, $v_d = V_{ds}^{'}/k T$, we find that for maintaining the same carrier density, one must meet the condition $q n_{s} = \frac{1}{2} C_{q} V_{th} \ln (1 + e^{\eta_{s} } ) (1 + e^{\eta_{s} - v_{d} }) $, which yields

\begin{equation}
\eta_{s} = \ln [ \sqrt{ (1 + e^{v_d} )^2 + 4 e^{v_d} ( e^{ \frac{2 q n_s}{C_q V_{th} } } - 1 )   } - (1 + e^{v_d} ) ] - \ln[2]
\label{etas}
\end{equation}
\begin{figure}[htb]
\centering
\includegraphics[width=60 mm]{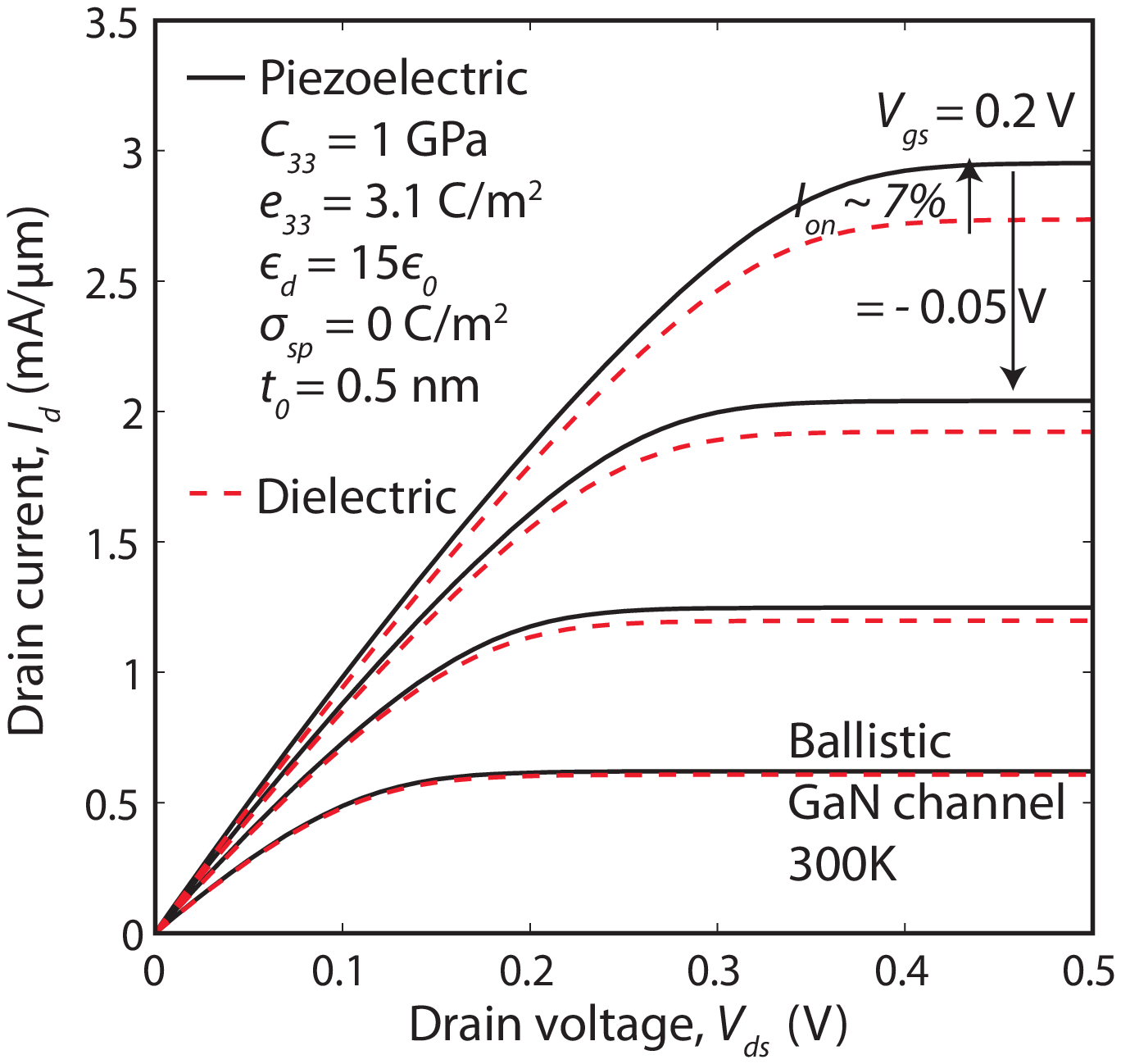}
\caption{ Output characteristics depicts the drain current $I_d$ versus $V_{ds}$ at gate bias voltage $V_{gs}=$ 0.2 V with a step of -0.05 V for GaN piezoFETs with piezoelectric (solid line) and dielectric barriers (dashed line).  A boost in the on-current is obtained for transistor with a compliant piezoelectric barrier than a dielectric one.}
\label{fig9}
\end{figure}

\begin{figure*}[htb]
\centering
\includegraphics[width=6 in]{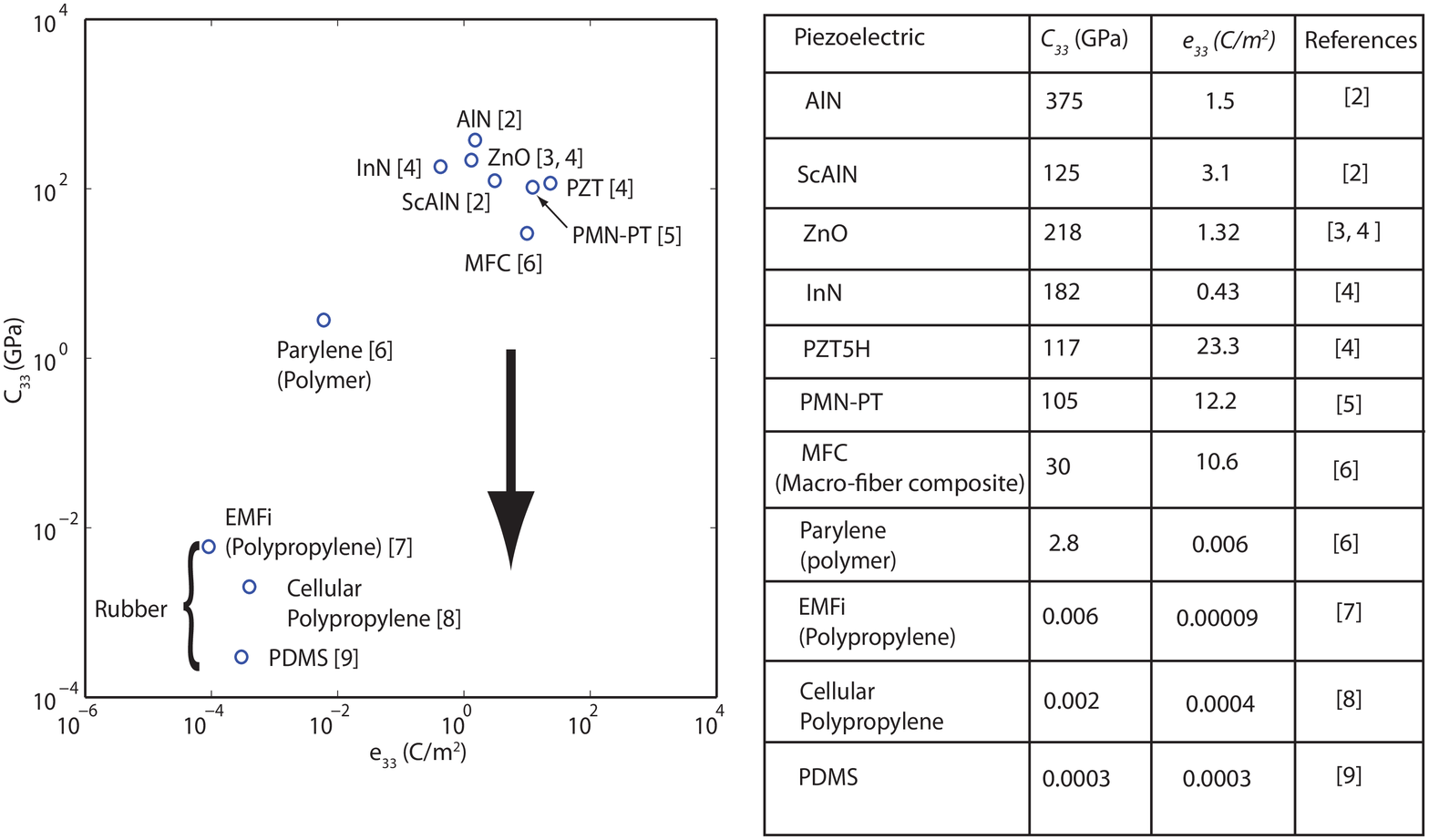}
\caption{ Piezoelectric Materials with $C_{33}$ and $e_{33}$ coefficients.  Table for different $C_{33}$ and $e_{33}$ coefficients of various piezoelectrics with corresponding references.  $C_{33}$ and $e_{33}$ values are obtained by literature review from listed references.  $e_{33}$ is extracted from a matrix relation of $[e]  = [C] *[d]$ where $[e]$ (unit: C/m$^2$) is the piezoelectric coefficient matrix, $[C]$ (unit: N/m$^2$) is the elastic stiffness matrix, [d] (unit: C/N) is the piezoelectric coupling matrix.  These matrices have sparsity pattern depends on the symmetry of the crystal.  However, here in order to get an estimation of the magnitude of $e_{33}$, we use a simpler relation $e_{33}$ (C/m$^2$) $=$ $C_{33}$ (N/m$^2$) $*$ $d_{33}$ (C/N), in cases where $e_{33}$ is not quoted in references.  The arrow shows the desired piezoelectric materials to be used for gate barriers of steep transistors.}
\label{chart}
\end{figure*}
Note that this is of a similar form as the seminal result on ballistic transistors by Natori et al \cite{NatoriJAP_94}, but the factor $n_{s}$ inside the square root must be obtained from Eq. \ref{piezoSCcharge} to account for the electromechanical coupling self-consistently.  By summing over the group velocities of the $k-$states, the net current per unit width of the ballistic piezoFET is then given by the same expression as in Natori \cite{NatoriJAP_94}:

\begin{equation}
J = J_{0} \left( F_{ \frac{1}{2} } ( \eta_s )  - F_{ \frac{1}{2} } ( \eta_s - \eta_d ) \right)
\label{ballisticcurrent}
\end{equation}

where $J_0 = \frac{ q g_s g_v \sqrt{2 m^{\star}} (kT)^{\frac{3}{2}}  }{ 2 \pi \hbar^2  }$, and $F_{\frac{1}{2}}$ is the Fermi-Dirac integral of order $1/2$.

The above expressions provide the complete electrical characteristics of a ballistic piezoFET at any temperature in a compact model.  We can obtain $n_s$ at any gate voltage $V_{gs}$ fully accounting for the electromechanical coupling by solving Eq. \ref{piezoSCcharge}.  We then find $\eta_s$ at any given $V_{gs}$ and $V_{ds}$ using Eq. \ref{etas}, and finally find the drain current per unit width using Eq. \ref{ballisticcurrent}.  Fig. \ref{fig9} shows the output characteristics depicting $I_d - V_{ds}$ at different $V_{gs}$ for GaN channel transistors with a compliant piezoelectric (solid lines) and passive dielectric (dashed lines) barriers.

Fig. \ref{chart} describes different piezoelectric materials with $C_{33}$ and $e_{33}$ coefficients obtained by literature review from Refs. \cite{Tasna´diPRL_10,Umeda,Wilson,CaoJAP04,Ramadan,Kressmanna,Heywang,Wang}.  Compliant piezoelectric materials with parameter space such as lower $C_{33}$ (higher compliance), and higher $e_{33}$ (higher piezoelectricity), shown by the arrow in Fig. \ref{chart} are useful for gate barriers of proposed steep transistors.
\section{Energy landscape for piezoelectric-semiconductor stack: Computational Details}
\subsection{Definition of Free-Energy}
\begin{figure}[t]
\centering
\includegraphics[width=40 mm]{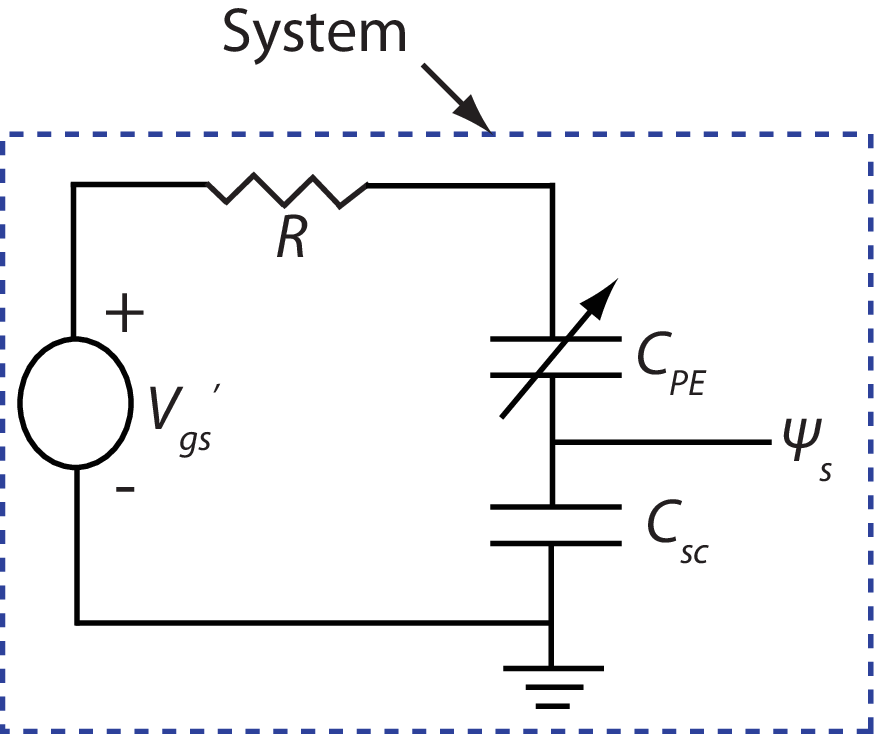}
\caption{ Schematic of a system shows a piezoelectric-semiconductor stack, represented by a series combination of piezoelectric capacitance $C_{PE}$ and semiconductor capacitance $C_{sc}$ connected to $V_{gs}^{'}$ via a resistor $R$. $\psi_s$ is the channel surface potential at the insulator-semiconductor interface.}
\label{circuit}
\end{figure}

Consider the circuit in Fig. \ref{circuit}, showing a piezoelectric-semiconductor stack connected to a voltage source via a resistor $R$ (units: $\Omega$cm${^2}$). Both the piezoelectric insulator and semiconductor have non-linear charge ($\sigma_{m}$) - voltage ($V$) characteristics, denoted say by $V=V_{ins} = f_1( \sigma_{m})$ and $\psi_s = f_2( \sigma_{m} )$ respectively.  A systematic method of studying the behavior of circuits with non-linear elements uses the Euler-Lagrange equations \cite{CherryPhilMag_1951}:
\begin{eqnarray}
\label{eq_Euler}
\mathcal{L}( \sigma_{m}, \dot{\sigma}_{m}  ) & =& T' - U \\
U( \sigma_{m} ) & =& \int_0^{\sigma_{m}} f_1( \tilde{\sigma}_{m} ) d \tilde{\sigma}_{m} \nonumber\\+
               \int_0^{\sigma_{m}} f_2( \tilde{\sigma}_{m}  ) d \tilde{\sigma}_{m}\\
\frac{ d \sigma_{m} }{dt} \Big( \frac{ \partial \mathcal{L} } { \partial \dot{\sigma}_{m} } \Big)  - \frac{ \partial \mathcal{L} } { \partial {\sigma_{m}}} & =& V_{gs}^{'} - R \dot{ \sigma}_{m},
\end{eqnarray}
where $\mathcal{L}( \sigma_{m}, \dot{\sigma}_{m}  )$ is the Lagrangian, $T'( \dot{\sigma}_{m} )$ is the magnetic co-energy (in any inductors) and $U( \dot{\sigma}_{m} )$ is the potential energy in the capacitors in the circuit.  For the circuit in Fig. \ref{circuit}, $T' = 0$, and we have
\begin{equation}
f_1(\sigma_{m}) + f_2(\sigma_{m}) = V_{gs}^{'} - R \dot{\sigma}_{m},
\end{equation}
which under equilibrium ($\dot{\sigma}_{m} = 0$) gives $V_{gs}' = f_1(\sigma_{m_0}) + f_2(\sigma_{m_0})$.  To determine whether $\sigma_{m_0}$ is a point of stable/unstable equilibrium, we write $\sigma_{m} = \sigma_{m_0} + \delta \sigma_{m_0}$ for a small perturbation $\delta \sigma_{m_0}$, expand the Lagrangian as a Taylor series (upto second order) about $\sigma_{m_0}$, and analyze whether the perturbation grows or decays with time:
\begin{eqnarray}
\dot{ \delta\sigma_{m} } & = - 2 \frac{f_1'(\sigma_{m_0}) + f_2'(\sigma_{m_0})}{R}  \delta\sigma_{m} = - \delta\sigma_{m} / \tau \\
\delta\sigma_{m}(t)  & = \delta\sigma_{m_0} \exp( - t/ \tau ).
\end{eqnarray}
where $f'(\sigma_m)$ is the derivative of $f(\sigma_m)$ w.r.t $\sigma_m$.
The system is stable to perturbations if $\delta\sigma_{m}(t) \to 0$ as $t \to 0$, i.e. if $f_1'(\sigma_{m_0}) + f_2'(\sigma_{m_0}) > 0$. The above discussion motivates the definition of the free-energy $G( \sigma_{m} )$:
\begin{equation}
G( \sigma_{m}, V_{gs}^{'} ) = \int_0^{\sigma_{m}} \left[ f_1(\tilde{\sigma}_{m}) + f_2(\tilde{\sigma}_{m}) \right] d\tilde{\sigma}_{m} - \sigma_{m} V_{gs}^{'}.
\label{freeenergy}
\end{equation}

From the carrier density expression in a semiconductor channel, $\psi_s$ can be expressed as $\psi_s = f_2(\tilde{\sigma_{m}})= V_{th}\ln\left(e^{\frac{\tilde{\sigma_{m}}}{C_{sc}V_{th}}}-1\right)$.  Using the expressions of $\psi_s$ and $V_{ins} = f_1(\tilde{\sigma_{m}})$ in Eq. \ref{freeenergy}, the free energy $G$ (unit: J/m$^2$) is found to be

\begin{eqnarray}
 G= \alpha_5 \sigma^{5}_{m} + \alpha_4 \sigma^{4}_{m} + \alpha_3 \sigma^{3}_{m} + \alpha_2 \sigma^{2}_{m} \nonumber\\+ V_{th} \int_{0}^{\sigma_{m}} \ln\left(e^{\frac{\tilde{\sigma}_{m}}{C_{sc}V_{th}}}-1\right)d\tilde{\sigma}_{m} - \sigma_{m}V_{gs}^{'},
\label{freeenergy1}
\end{eqnarray}

where $\alpha_5 = \frac{ e_{33} }{5C_{0} (\epsilon_d C_{33})^2 }$, $\alpha_4 = \frac{1}{4C_{0} \epsilon_d C_{33}}$, $\alpha_3 = \frac{ ( \sigma_{sp} - e_{33} ) }{3C_{0} \epsilon_d C_{33} }$, $\alpha_2 = \frac{ 1}{2C_{0} }$ are material and geometric constants of the problem.  Note the internal energy is a non-linear function of the sheet charge $\sigma_m$, and is a linear function of applied gate bias voltage $V_{gs}$.  Then, all points of equilibrium satisfy ${\partial G} / { \partial\sigma_{m}} = 0$, and further, points of {\em stable} equilibrium satisfy ${\partial^2 G} / {\partial \sigma_{m}^2} > 0 $.

\subsection{Self-consistent solution}
We make use of the Euler-Lagrange equation to calculate the $I-V$ characteristics of the PiezoFET.  The basic structure of the self-consistent algorithm involves an outer loop that determines the voltage drop across the intrinsic FET (i.e eliminating the voltage drops across the contacts). Inside this loop, we require the charge $n_s$ for the intrinsic gate voltage. To do this, we solve the Euler-Lagrange equation (using the  implicit Euler method for about $5\tau$) to determine a good guess of charge $n_{s, guess}$ at the present voltage, starting from the value of $n_{s0}$ obtained at a previous voltage. $n_{s, guess}$ is then used  in a fixed point iteration scheme to determine $n_s$, from which the current is finally calculated.  We have not encountered prior use of this Euler-Lagrange method for the calculation of hysteretic characteristics of electron device systems in an extensive literature search, and intend to publish the detailed procedure in a follow-up report.

\end{document}